\documentclass[11pt]{article}
\usepackage[T1]{fontenc}
\usepackage[latin1]{inputenc}
\usepackage[dvips]{epsfig}
\usepackage{graphicx}
\usepackage[english]{babel}
\usepackage{amsmath}
\usepackage{amssymb}
\usepackage{amsfonts}
\usepackage{color}
\DeclareMathOperator{\sech}{sech}
\DeclareMathOperator{\LambertW}{LambertW}
\DeclareMathOperator{\arcsinh}{arcsinh}
\begin{document}

\title {\large{ \bf {{FRW}} string cosmological solutions via Hojman symmetry}}

        \vspace{3mm}

        \author {  \small{  F. Darabi}\hspace{-1mm}{ \footnote{ E-mail: f.darabi@azaruniv.ac.ir
(Corresponding author.)}} , \small{   M. Golmohammadi}\hspace{-1mm}{
                        \footnote{ E-mail: golmohammadi@azaruniv.ac.ir
                                }}, \small{  A. Rezaei-Aghdam }\hspace{-1mm}{\footnote{E-mail: rezaei-a@azaruniv.ac.ir}} \\ \\
                {\small{\em Department of Physics,  Azarbaijan Shahid Madani University}},
                {\small{\em   53714-161, Tabriz, Iran  }}}
        \maketitle

\begin{abstract}
 In this paper, we  find  exact string cosmological solutions for FRW cosmology, by using Hojman symmetry approach.
The string cosmology under consideration includes a scalar field $\psi(t)$ with the potential $W(\psi)$, and     a totally antisymmetric field strength $H_{\mu\nu\rho}$ which is
 specifically defined in terms of the scale factor $a(t)$. We show that for this string
cosmology,  Hojman conserved quantities exist    using which new
exact solutions  for
the  scale factor and the scalar field are obtained for  specific potentials $W(\psi)$ with some free parameters. The presence
of these parameters, together with those of arising from Hojman symmetry,  is an important advantage using which
one can construct the  solutions $a(t)$ and $\psi(t)$ with variety of cosmological behaviors.     
\end{abstract}
%%%%%%%%%%%%%%%%%%%%%%

%%%%%%%%%%%%%%%%%%%%%%

\section{Introduction}
 String theory has major objectives,  the most recent one of which  is the development
of a well-defined cosmological framework to upgrade the conventional  cosmology near the Planck  scale. This is  the scale at which
most of the major cosmological problems arise and hence one needs a very
 sophisticated insight to resolve them. The cosmology at this era is offered by string theory as ``String Cosmology'' which is supposed to provide a sufficient
understanding of the very early universe  and a subsequent graceful exit towards the conventional ``Hot Big Bang cosmology''.

{According to the recent observations,  the
dynamics of early universe might have been profoundly affected by the presence of
spatial anisotropies near the Planck scale \cite{Bucher}. Motivated by this argument, in the study of string cosmology, some people
have relaxed the requirement of spatial isotropy
over the 4D spatially homogeneous
 cosmologies   and  focused on the
spatially homogeneous but not necessarily
isotropic 4D string backgrounds \cite{Fradkin}-\cite{Casprini}. This class of cosmologies is considered
as the string cosmology counterpart of  the vacuum Bianchi-type models \cite{Callum}-\cite{Kolb}. Vacuum Bianchi-type models  involves 4D spatially homogeneous, but not isotropic,
spacetimes
which satisfy at least the lowest-order string beta-function equations \cite{Batakis}. These models generalize all possible FRW cosmological models and provide the best models available for understanding
the impacts of anisotropy on the dynamics of  early universe.
Although the above mentioned arguments favor the Bianchi-type cosmological
models  involving 4D spatially homogeneous and non-isotropic
spacetimes, however we can still trust in homogeneous and isotropic
FRW cosmological models to study the early universe string cosmology \cite{ven}.
Therefore, in this paper, we use FRW homogeneous and isotropic
spacetime for the sake of simplicity. This choice is  mainly motivated by the fact that the  formulation of Hojman symmetry  (see below) favors the least numbers of dynamical variables to
achieve the straightforward results. }

Symmetries and the corresponding conserved quantities,
in general, help to simplify the dynamics and
give rise to exact solutions for physical systems under consideration. Noether symmetry  is
a well known example  which is  widely used in  different aspects of classical and quantum
field theory, as well as in general relativity and black hole physics.
Noether symmetry
approach has been received great amount of attention in the context of  cosmology \cite{aghdam}-\cite{Aslam}. Imposing Noether
symmetry on the point-like Lagrangians, associated
to the equations of motion of a cosmological model, allows one to find  the first integrals of the equations
of motion.
Recently, an alternative  approach to  the Noether
symmetry approach has been received attention, so called
Hojman symmetry approach, which can provide us with a  method to find new exact solutions \cite{Hoj}. Unlike the Noether symmetry
approach, in the Hojman
symmetry approach Lagrangian and Hamiltonian functions are not required
to find the exact solutions, rather the symmetry
vectors and the corresponding conserved charges
are obtained by using  the equations of motion, without referring to Lagrangian or  Hamiltonian.
It turns out that these two approaches
may give rise to different conserved quantities
as well as different exact solutions.   Recently, parallel to the Noether symmetry, the Hojman symmetry has also been extensively used to study some models of extended theories of gravity and cosmology \cite{Myrzakul}-\cite{Hoj5}.  In this paper, we study and apply the Hojman symmetry for some {{FRW}} string cosmological models, to obtain  new conserved charges and exact solutions.

The organization of the paper is following. In Sec. 2, we briefly review the main points of the String Cosmology.  In Sec. 3, we have a summery look at the Hojman symmetry approach and in Sec. 4, we use Hojman symmetry method in {{FRW}}  sting cosmological model to find new solutions.
In section 5, we investigate the possible inflationary behavior of the solutions
by implementing the slow roll approximations. The paper ends with a conclusion, in section 6.
\section{Preliminaries in String Cosmology }
Cosmology is a framework to describe the evolution of the Universe, starting
from its
very beginning at  high energy regime of Planck scale, namely quantum gravity
era.
On the other hand,
string theory is considered as  the only known consistent theory of quantum gravity, at Planck energy scale. Therefore, it is appealing to study
 the very early Universe in the context of string theory, and conversely, the
very early Universe seems to be a very natural place to establish some predictions for string models (for more details see\cite{Nastase}). In this line of study,
string-dialton cosmology has come out from the low energy limit of superstring theory, and  some exact solutions  has been obtained
in homogeneous isotropic backgrounds \cite{Capoziello1}-\cite{Capoziello2}.  Moreover, it has been shown that the gravity may couple, in addition to scalar field, to an antisymmetric second rank Kalb-Ramon tensor, so called $B$-field. In this model, at low energy limit with a very weak coupling, one may describe a string dominated universe  by the tree level effective action including only massless modes, namely tensor modes (graviton) and scalar modes (dilaton), the exact solutions of which for homogeneous and anisotropic  string cosmological models    are studied in 4D  \cite{Batakis} and 5D  \cite{aghdam}.
However, at high energy limit with strong coupling, the
tensor modes and scalar modes  experience   considerable growths \cite{naderi0}. Therefore,  true string cosmological description of the universe, at high energy regime of Planck scale, necessitates the use of perturbative corrections
in string cosmology, {including the stringy type $\alpha'$-expansion and
the quantum loop expansion in string coupling \cite{naderi1}},  \cite{naderi2},  \cite{naderi3},  \cite{naderi4}.

  Here, we will assume the 4-dimensional spacetime curvatures below the string-Planck scale and describe
the evolution of the universe  by the following effective action \cite{kikawa}
\begin{equation}\label{ef}
S_{eff}=\int\mathrm{d}^4x\,\sqrt{-g}e^\psi\left( R-\frac{1}{12}H_{\mu\nu\rho}H^{\mu\nu\rho}+\partial_\mu\psi\partial\,^\mu\psi-W(\psi)\right) ,
\end{equation}
where $g_{\mu\nu}$ is  the  background metric, non-minimally coupled to a scalar field $\psi$   with the potential $W(\psi)$, and $H_{\mu\nu\rho}$  is a totally antisymmetric field strength  which is
 defined in terms of the antisymmetric Kalb-Ramon field $B_{\mu\nu}$ as
\begin{equation}\label{bmunu}
  H_{\mu\nu\rho}=\partial_\rho B_{\mu\nu}+\partial_\nu B_{\rho\mu}+\partial_\mu B_{\nu\rho}.
\end{equation}
The field equations can be derived as follows (see \cite{Fradkin} and \cite{Gross})
  \begin{align}\label{bequations}
   R-\frac{1}{12}H_{\mu\nu\rho}H^{\mu\nu\rho}-\partial_\mu\psi\partial^\mu\psi-2\nabla_\mu\nabla^\mu\psi-W(\psi)-\frac{\mathrm{d}W}{\mathrm{d}\psi} &=0, \\
    R_{\mu\nu}-\frac{1}{4}H^2_{\mu\nu}-\nabla_\mu\nabla_\nu\psi-\frac{1}{2}g_{\mu\nu}\frac{\mathrm{d}W}{\mathrm{d}\psi}&=0,\\
    \nabla_\mu\left( e^\psi H_{\mu\nu\rho}\right) &=0.
  \end{align}
We take the flat Friedmann-Robertson-Walker (FRW) background metric
\begin{equation}\label{metric}
dS^2=-dt^2+a(t)^2(dx^2+dy^2+dz^2),
\end{equation}
where $a(t)$ is the scale factor.
In the case of homogeneous cosmological backgrounds, $\psi$ should also be a monotonic
function of  time $t$. Also, according to \cite{Batakis}
and with no loss of generality, we may consider the 3-form  
\begin{equation}\label{H}
  H=a^2(t)\,\mathrm{d}t\wedge\mathrm{d}x\wedge\mathrm{d}y,
\end{equation}
for the $H$ field which is satisfied by the equation (2).

\section{Brief review on Hojman conservation method}
Hojman symmetry was proposed in 1992 \cite{Hoj}, a brief review of which is as follows. Consider a set of second-order differential equations
 \begin{eqnarray}\label{force}
        \ddot{q}^i=F^{i}(q^j,\dot{q}^j,t),\qquad i,j=1,\,2,\,...\,n
\end{eqnarray}
where $q^{i}$ denotes the coordinates, $F^{i}$ denotes the forces, and a dot denotes  derivative with respect to time $t$. If this equation
has a symmetry vector $X^i=X^i(q^j,\dot{q}^j,t)$, then it has to satisfy the following equation
(\cite{Hoj1},  \cite{Hoj2})
   \begin{equation}\label{sym.vector}
     \frac{\mathrm{d}^2X^i}{\mathrm{d}t^2} - \frac{\partial F^i}{\partial q^j}\, X^j - \frac{\partial F^i}{\partial \dot{q}^j} \, \frac{\mathrm{d}X^j}{\mathrm{d}t} =\,0,
   \end{equation}\\
   where\\
   \begin{equation}\label{d/dt}
      \frac{ \mathrm{d}}{\mathrm{d}t}\,=\,\frac{\partial}{\partial t}+\dot{q}^i \, \frac{\partial}{\partial q^j} \, +{F}^i \, \frac{\partial}{\partial \dot q^i}.
   \end{equation}\\
The symmetry vector $X^i$ has the property that  under the infinitesimal transformation
\begin{equation}\label{infi. trans}
\nonumber
  \hat{q}^i=q^i+\epsilon X^i \left( q^j, \dot{q}^j, t \right),
\end{equation}
the solutions $q^i$ of Eq.(\ref{force}) are mapped into the solutions $\hat{q}^i$
of the same equations (up to $\epsilon^2$ terms) \cite{Hoj3}, \cite{Hoj4}.
With this property of the symmetry vector $X^i$, the Hojman conserved quantities are defined through the following theorem \cite{Hoj5}:\\
\\ {\bf Theorem}:\\
    1. If the force $F^i$ satisfies
the following equation   
\begin{equation}\label{cons. quantity0}
      \nonumber
      \frac{\partial F^i}{\partial \dot{q}^i} =0 \, ,
   \end{equation}
 then

  \begin{equation}
    Q=\frac{\partial X^i}{\partial q^i}+ \frac{\partial }{\partial \dot{q}^i} \left( \frac{\mathrm{d}X^i}{\mathrm{d}t}\right),
  \end{equation}
is a conserved quantity for Eq. (\ref{force}), i.e. $\frac{\mathrm{d}Q}{\mathrm{d}t}=0$,\\
\\
 2. and if $F^i$ satisfies

   \begin{equation}
   \nonumber
        \frac{\partial F^i}{\partial \dot{q}^i} =-\frac{\mathrm{d}}{\mathrm{d}t}\ln \gamma,
   \end{equation}
then
  \begin{equation}\label{cons. quantity1}
   Q =\frac{1}{\gamma} \frac{\partial (\gamma X^i)}{\partial q^i}+ \frac{\partial }{\partial \dot{q}^i} \left( \frac{\mathrm{d}X^i}{\mathrm{d}t}\right),
  \end{equation}
is a conserved quantity, where $\gamma$ is merely a function of $q^i$.\\

  \section{{{FRW}}  \textit{} String Cosmology via Hojman symmetry}
  In this section, we will apply the Hojman symmetry approach on the  string cosmological system introduced in section 2. Our starting point is the {{scalar-tensor}} Lagrangian density
  
  \begin{equation}
  \mathcal{L}=\sqrt{-g}e^\psi\left( R-\frac{1}{12}H_{\mu\nu\rho}H^{\mu\nu\rho}+\partial_\mu\psi\partial\,^\mu\psi-W(\psi)\right).
  \end{equation}
By considering the transformations \cite{capponoether}

\begin{equation}\label{Conf}
\varphi(\psi)={{2}} e^{\frac{1}{2}\psi},\qquad \mathcal{F}(\varphi)
={\frac{1}{4}}(\varphi)^2,\qquad V(\varphi)=e^\psi W(\psi),
\end{equation}
we obtain the equivalent Lagrangian density
\begin{equation}\label{LD}
   \mathcal{L}=\sqrt{-g}\left(\mathcal{F}(\varphi)R-\frac{1}{12}\mathcal{F}(\varphi)H_{\mu\nu\rho}H^{\mu\nu\rho}+\partial_\mu\varphi\,\partial\, ^\mu\varphi-V(\varphi)\right).
\end{equation}
 Using the FRW metric (\ref{metric}) and the 3-form (\ref{H}), the point-like
Lagrangian density (\ref{LD}) reads
as
\begin{equation}{\label{lagrange}}
  \mathcal{L}  =-6 \mathcal{F}(\varphi)a\dot{a}^2-6 \mathcal{F}'(\varphi)a^2\dot{a}\dot{\varphi}-a^3\dot{\varphi}^2-a^3Z(\varphi),
\end{equation}
where $Z(\varphi)= V(\varphi)-\frac{1}{2}\mathcal{F}(\varphi)$. The Euler-Lagrange equations corresponding to (\ref{lagrange})  are obtained

\begin{align}
   &4\frac{\ddot{a}}{a}+2\frac{\dot{a}^2}{a^2}+4\frac{\mathcal{F}'}{\mathcal{F}}\frac{\dot{a}}{a}\dot{\varphi}+2\frac{\mathcal{F}'}{\mathcal{F}}\ddot{\varphi}+\left(2\frac{\mathcal{F}''}{\mathcal{F}}-\frac{1}{\mathcal{F}}\right)\dot{\varphi}^2-\frac{1}{\mathcal{F}}Z(\varphi)=0,\\
   &\ddot{\varphi}+3\mathcal{F}'\frac{\dot{a}^2}{a^2}+3\mathcal{F}'\frac{\ddot{a}}{a}+3\frac{\dot{a}}{a}\dot{\varphi}-\frac{1}{2}Z'(\varphi)=0,
\end{align}
subject to the zero energy constraint
\begin{equation}\label{0,0 function}
  -6\mathcal{F}a\dot{a}^2-6\mathcal{F}'a^2\dot{a}\,\dot{\varphi}-a^3\dot{\varphi}^2+a^3Z(\varphi)=0,
\end{equation}
where $'$ denotes the differentiation $h^{'}(y)=dh/dy$.

{{In the cosmological study, there are  two well known frameworks
 as  {\it Einstein frame}
 and  {\it Jordan frame}. The former case 
is the frame with a Lagrangian  having minimally coupled  terms,
whereas the later
case is the frame with a Lagrangian density having non-minimally coupled  terms. The Lagrangian (\ref{0,0 function}) is of the  later  type because of the  first two non-minimally
coupled  terms. For
any non-minimally coupled Lagrangian,  a unique
minimally coupled Lagrangian may be associated  by
establishing a  relation between a suitable combination of the coupling (here $\mathcal{F}$) and the potential (here $Z(\varphi)$) in the non-minimal Lagrangian, in one hand, and the
potential in the minimal Lagrangian, on the other hand. Therefore, using this correspondence,
we may find the solutions for the
non-minimally coupled model in Jordan frame through finding the corresponding solutions in the minimally coupled model in Einstein
frame and transforming them back into  Jordan frame using
the inverse conformal transformations. We will use this strategy, similar
to \cite{Hoj2}, in the sense that by obtaining the solutions, through the Hojman Theorem in
 Einstein frame,  it
is possible to find the corresponding solutions in  Jordan frame.
}}

Now, it is possible to use the following {\it conformal} transformations \cite{Hoj} and \cite{allemandi}
   \begin{align}\label{cft}
       \bar{a}=\sqrt{2\mathcal{F}}\,a~, ~~~~~
       \mathrm{d}\bar{t}=\sqrt{2\mathcal{F}}\,\mathrm{d}t~, ~~~~~
      \frac{\mathrm{d}{\bar{\varphi}}}{\mathrm{d}t}=\sqrt{\frac{3\mathcal{F}'^2-{2}\mathcal{F}}{2\mathcal{F}^2}}\,\frac{\mathrm{d}\varphi}{\mathrm{d}t},
   \end{align}
under which the Lagrangian (\ref{lagrange}) takes on the  form,
corresponding to  a minimally coupled scalar field, as follows
\begin{equation}\label{cftl}
  \mathcal{\bar L}={{-}}3\bar{a}\dot{\bar{a}}^2{{+\frac{1}{2}}}\bar{a}^3\dot{\bar{\varphi}}^2-\bar{a}^3\bar{Z}(\bar{\varphi}),
\end{equation}
where $\bar{Z}(\bar{\varphi})=\frac{Z(\varphi)}{4\mathcal{F}^2}$ and $\dot{\bar
}$ denotes derivative with respect to $\bar{t}$.  
The Euler-Lagrange equations corresponding to (\ref{cftl})  are also obtained
as
 \begin{equation}\label{abar}
    2\frac{\ddot{\bar{a}}}{\bar{a}}+\frac{\dot{\bar{a}}^2}{\bar{a}^2}+{{\frac{1}{2}}}\dot{\bar{\varphi}}^2-\bar{Z}(\bar{\varphi})=0,
   \end{equation}
      \begin{equation}\label{phibar}
    {\ddot{\bar{\varphi}}+3\frac{\dot{\bar{a}}}{\bar{a}}\dot{\bar{\varphi}}+\bar{Z}'=0},
    \end{equation}
subject to the zero energy constraint
\begin{equation}\label{ZEC2}
 -3\frac{\dot{\bar{a}}^2}{\bar{a}^2}+{{\frac{1}{2}}}\dot{\bar{\varphi}}^2+\bar{Z}(\bar{\varphi})=0,
\end{equation}
where $'$ denotes the differentiation $\bar h^{'}(\bar y)=d\bar h/d\bar y$.
Combining (\ref{abar}) and (\ref{phibar}) with (\ref{ZEC2}) leads to
 \begin{align}\label{onedynamical}
      &2\left(\frac{\ddot{\bar{a}}}{\bar{a}}-\frac{\dot{\bar{a}}^2}{\bar{a}^2}\right)+{{\dot{\bar{\varphi}}^2}}=0,\\
         &{\frac{\bar{Z}'}{\bar{Z}}=-\frac{\ddot{\bar{\varphi}}+3\frac{\dot{\bar{a}}}{\bar{a}}\dot{\bar{\varphi}}}{3\frac{\dot{\bar{a}}^2}{\bar{a}^2}-\frac{1}{2}\dot{\bar{\varphi}}^2}},
 \end{align}
respectively. The property that $\bar a(t)$ and $\bar{\varphi}(\bar{t})$
are invertible functions of time $t$, helps us to simplify the dynamics by
reducing  two above dynamical equations to  one dynamical equation. In doing so, we define $\bar x=\ln\bar{a}$,  $\bar{\varphi}(\bar{t})=\bar{\varphi}(\bar x(\bar{t}))$,    and use them in Eqs. (\ref{onedynamical}) and {{(26)}} to obtain the following  dynamical equations
   \begin{equation}\label{ddotx}
     \ddot{\bar{x}}=-{\frac{1}{2}}\bar{\varphi}'^2(\bar x)\dot{\bar x}^2,
   \end{equation}
   and
   \begin{equation}\label{zdotx}
     \frac{\bar{Z}'}{\bar{Z}}={{-\bar{\varphi}'(\bar x)+\frac{2}{\bar{\varphi}'^2(\bar x)-6}\bar{\varphi}''(\bar x)}}.
   \end{equation}

Now, these equations are in their most suitable forms to be studied in
the context of Hojman symmetry approach. Assuming the one dimensional vector $X(\bar x,\dot{\bar x})$, independent of explicit time $t$, the equation of
symmetry vector $X$ is obtained as
\begin{equation}\label{Equation}
 \left( \frac{\partial\,^2X}{\partial \bar x^2}+f'(\bar x)X+f(\bar x)\frac{\partial X}{\partial \bar x}\right)+\dot{\bar x}^2f^2(\bar x)\frac{\partial\,^2X}{\partial \dot{\bar x}^2}-\dot{\bar x}\left(2f(\bar x)\frac{\partial\,^2X}{\partial \bar x\partial \dot{\bar x}}+f'(\bar x)\frac{\partial X}{\partial \dot{\bar x}}\right)=0,
\end{equation}
where
\begin{equation}\label{f(x)}
f(\bar x)={{\frac{1}{2}}}\bar{\varphi}'^2(\bar x).
\end{equation}

 From equation (\ref{ddotx}), we can recognize that $F(\bar x, \dot{\bar x})=-f(\bar x){\dot{\bar x}}^2$, thus
 \begin{equation}{\label{gamma}}
 \gamma(\bar x)=\gamma_0\, e^{2\int{f(\bar x)\, \mathrm{d}\bar x}},
 \end{equation}
 {{where $\gamma_0$ is an integration constant.}} 
 
 {{In general, the differential equation for vector $X$ is difficult to solve.
In order to proceed in solving  the equation (\ref{Equation}), we will limit ourselves to some  particular ansatz,  below, for vector $X$  proposed in the {{scalar-tensor}} framework \cite{Hoj1}. }}

\subsection{$X \sim X(\dot{\bar x})$}
  
  By considering $X \sim X(\dot{\bar x})$, the corresponding differential equation for symmetry vector $X$ according to (\ref{Equation}) takes the form  
  \begin{equation}
 \frac{\mathrm{d}^2 X}{\mathrm{d}\bar x^2}+f'(\bar x) X +f(\bar x) \frac{\mathrm{d}X}{\mathrm{d}\bar x}=0,
  \end{equation}
   or   \begin{equation}
   \frac{\mathrm{d}}{\mathrm{d}\bar x}\left( f(\bar x)X+\frac{\mathrm{d}X}{\mathrm{d}\bar x}\right)=0,
   \end{equation}
 which is
      nothing but $\frac{\mathrm{d}Q}{\mathrm{d}\bar x}=0$, according to Eq. (\ref{cons. quantity1}), so we will not consider this choice. For the choice $X=X(\dot{\bar x})$,  the equation of symmetry  vector $X$ reads as Euler equation
\begin{equation}\label{Equation1}
  f'(\bar x)X+\dot{\bar x}^2f^2(\bar x)\frac{\mathrm{d}\,^2X}{\mathrm{d} \dot{\bar x}^2}-\dot{\bar x}f'(\bar x)\frac{\mathrm{d} X}{\mathrm{d} \dot{\bar x}}=0,
\end{equation}
from which $X$ and $f(\bar x)$ are obtained respectively as
follows
\begin{equation}
  X=A_1\dot{\bar x}+A_2\dot{\bar x}^n,
\end{equation}
\begin{equation}
  f(\bar x)=-\frac{1}{n\bar x+f_0},
\end{equation}
where $A_1, A_2, f_0$ and $n$ are constant parameters, and the Hojman conserved quantity reads as
\begin{equation}
        Q=2f(\bar x) \dot{\bar x}^n-f(\bar x)n(n+1)\dot{\bar x}^n.
\end{equation}
 We can easily verify
that the symmetry vectors $X \sim \dot{\bar x}$ and $X \sim \dot{\bar x}^{-2}$ give rise to vanishing conserved charge, namely $Q=0$. Therefore, we may discard $n=1,-2$
cases.

If we define a new variable as $\bar y=-(n\bar x+f_0)>0$, then using (\ref{zdotx}) and (\ref{f(x)}), we obtain

\begin{equation}\label{1varphi}
  \bar{\varphi}=\bar\varphi_c\mp{{\frac{2\sqrt{2}}{n}}}\sqrt{\bar
  y},
\end{equation}
\begin{equation}\label{2varphi}
  \bar{Z}(\bar{\varphi})={{\lambda(\bar{\varphi}-\bar{\varphi_c})^\frac{4}{n}-\frac{8}{3n^2}\lambda(\bar{\varphi}-\bar{\varphi_c})^{\frac{4}{n}-2}}},
\end{equation}
{{where $\bar{\varphi_c}$ is a constant, ${{\lambda=3\bar{Z}_0(\frac{n^2}{8})^\frac{2}{n}}}$
and $\bar{Z}_0$ is a constant of integration.}}
On the other hand, using the equations (\ref{cons. quantity1}) and (\ref{gamma}), for the conserved quantity we have
\begin{equation}\label{Q-0}
Q_0=\frac{\dot{\bar x}^n}{n\bar x+f_0}.
\end{equation}
As was mentioned in \cite{Hoj1}, for $\dot{\bar x}^n>0$ and  $\dot{\bar x}^n<0$ one obtains  $Q_0<0$ and $Q_0>0$, respectively. On the other hand, since $\dot{\bar x}$ can be negative or positive, we can assume  $n$ to be an integer. Using $\bar y=-(n\bar x+f_0)$, we have
\begin{equation}
\dot{\bar y}^n=(-n)^n\bar y|Q_0|.
\end{equation}
The solution of this equation is obtained as
\begin{equation}\label{y}
\bar y(\bar t)=\left[\left( 1-\frac{1}{n}\right) \left(\bar y_0-n|Q_0|^{\frac{1}{n}}\bar
t\right)\right]  ^{\frac{n}{n-1}},
\end{equation}
where $\bar y_0$ is an integration constant. {{Considering (\ref{y}), (\ref{1varphi}), (\ref{2varphi}), and putting them in (\ref{phibar}) we find $\bar{Z}_0=|Q_0|^{\frac{2}{n}}$
 for which the solutions satisfy the field equations.}
}

{Using equations (\ref{1varphi}) and (\ref{y}), we can write the solutions $\bar{a}(\bar t)$ and $\bar{\phi}(\bar t)$ for the potential (\ref{2varphi}) as follows
\begin{equation}
        \bar{a}(\bar t)= e^{-\frac{f_0}{n}} \, e^{- \frac{1}{n}\left((1- \frac{1}{n})( y_0-n|Q_0|^{\frac{1}{n}}\bar t)\right)^ \frac{n}{n-1}},
\end{equation}
\begin{equation}
        \bar{\phi}(\bar t)=\bar{\phi_c} \pm \frac{2\sqrt{2}}{n} \left [  (1-\frac{1}{n})( y_0-n|Q_0|^{\frac{1}{n}}\bar t)\right]^ \frac{n}{2(n-1)},
\end{equation}
which  can be transformed, using the transformations (\ref{cft}), to
\begin{equation}\label{scalef.I}
          a(\tau)= \sqrt{2} e^{-\frac{f_0}{n}} \, e^ { \pm \frac{2}{n}((1- \frac{1}{n})\tau)^ \frac{n}{2(n-1)}}\, e^{- \frac{1}{n}((1- \frac{1}{n})\tau)^ \frac{n}{n-1}},
\end{equation}
\begin{equation}
        \phi(\tau)=e^{\mp\frac{2}{n}\left( (1-\frac{1}{n})\tau\right) ^{\frac{n}{2(n-1)}}},
\end{equation}
where $\tau=y_0-n|Q_0|^ \frac{1}{n}\,\bar t $\, .
Finally, by using the transformations (\ref{Conf}),  we obtain 
 \begin{equation}\label{dilaton.I}
    \psi(\tau)= 2\ln \frac{1}{2} \mp \frac{4}{n}\left [  (1-\frac{1}{n})\tau\right]^ \frac{n}{2(n-1)},
  \end{equation} 
   \begin{equation}\label{potential.I}
    W(\psi)=4\lambda e^\psi \left[(\sqrt{2}\ln 2+\frac{\sqrt{2}}{{2}}\psi)^\frac{4}{n}-\frac{8}{3n^2}((\sqrt{2}\ln 2+\frac{\sqrt{2}}{{2}}\psi)^{\frac{4}{n}-2}\right]+\frac{1}{2}\:.
  \end{equation}
Thus, if we consider $X \sim \dot{x}^n$, the generic potential $W(\psi)$ takes the form (\ref{potential.I}).
 In the following, we will consider more general cases where the symmetry vector $X$ depends on both $x$ and $\dot{x}$.}
  
{ {Because of variety of free parameters in the solutions, one may obtain different time dependent behaviors
for different choices of these parameters. Apart
from the integration constant $\bar{y}_0$, which can be absorbed in the time
reparametrization, the solutions
depend implicitly on the conserved charge
$Q_0$ as well as  the constant parameters, namely $f_0$ and $n$,  related to the conserved charge
$Q_0$ through Eq.(\ref{Q-0}).  For the sake of briefness, among all  solutions   obtained in this
paper, we will limit ourselves to discuss on the viable solutions relevant to the
present day cosmological models, including inflationary or late time accelerating behaviors.  }} 

{{ In Figure 1, the diagrams of $a(\bar t)$, $\psi(\bar t)$, and $q(\bar t)=-a\ddot a/\dot a^2$ (deceleration parameter) are depicted for the solutions (\ref{scalef.I}) and (\ref{dilaton.I}), with lower signs, for some typical values of the parameters $Q_0, {f_0}, {\bar y_0}, $ and $n$ (Here   dot $\dot {}$ denotes differentiation $d/d{\bar t}$). This solution describes a universe which   experiences an initial transient decelerating  expansion, followed by an 
eternal accelerating
de Sitter expansion.  }} 
\begin{figure}[ht]
  \centering
  \includegraphics[width=2in]{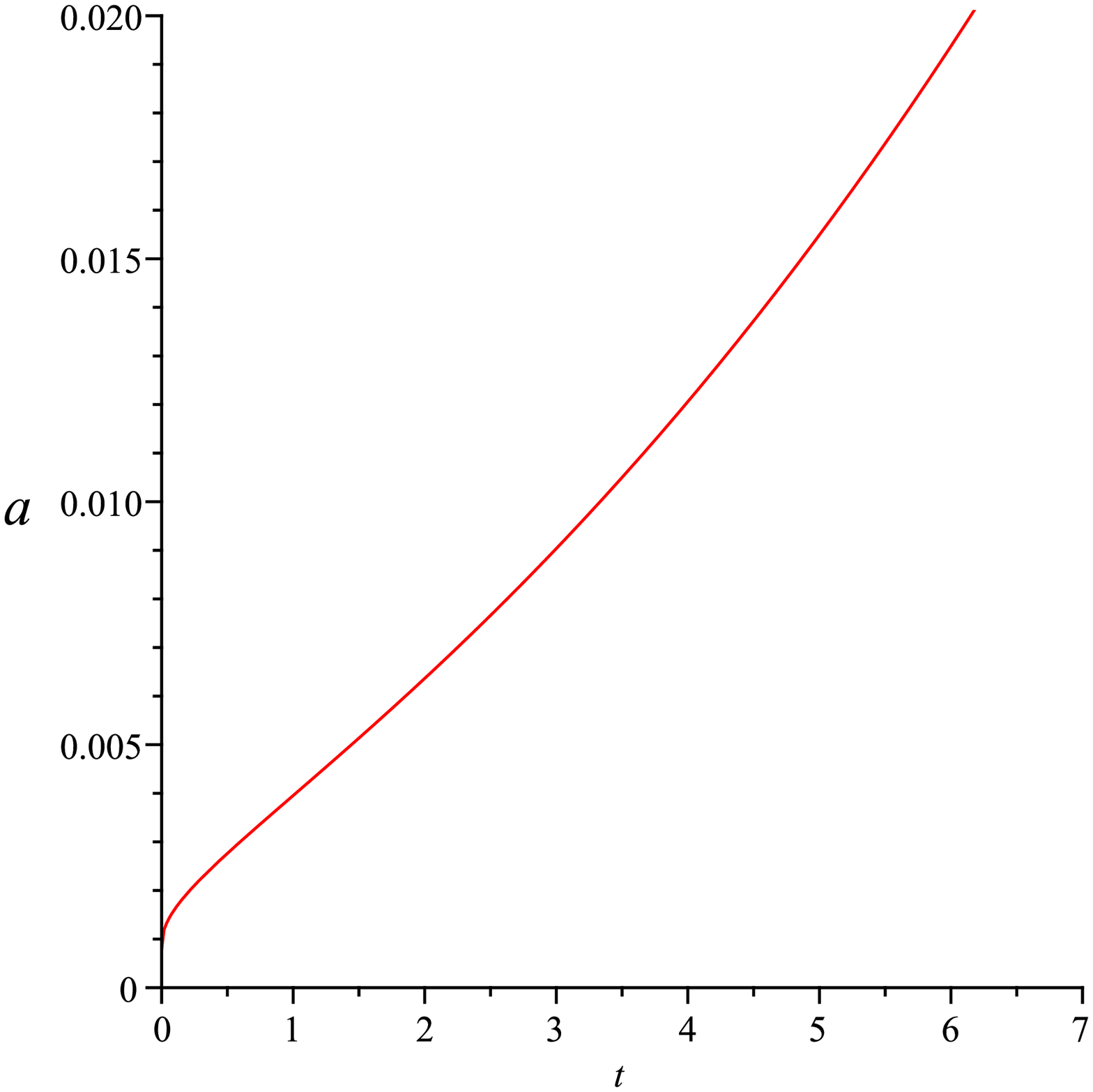}
   \includegraphics[width=2in]{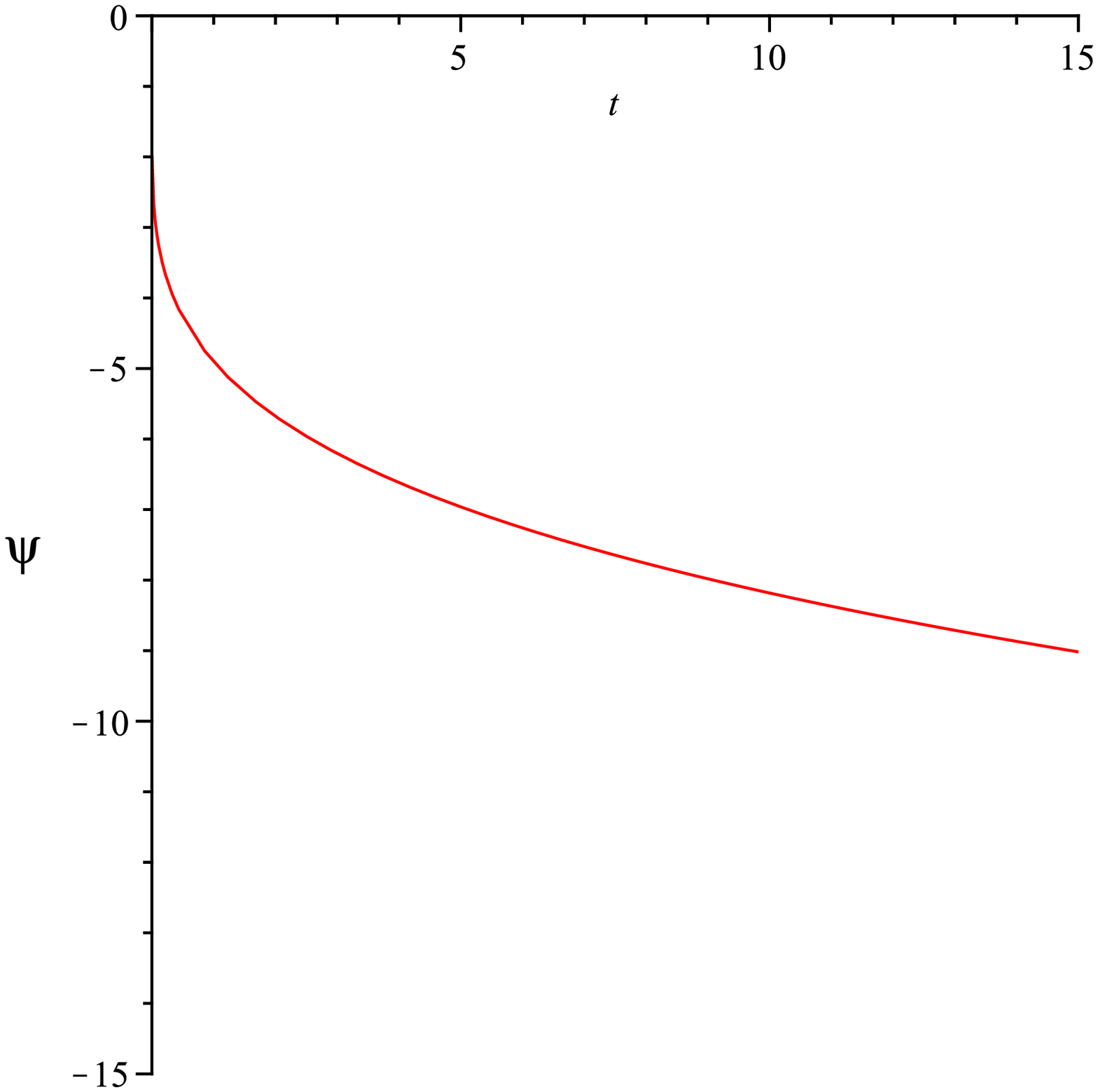}
    \includegraphics[width=2in]{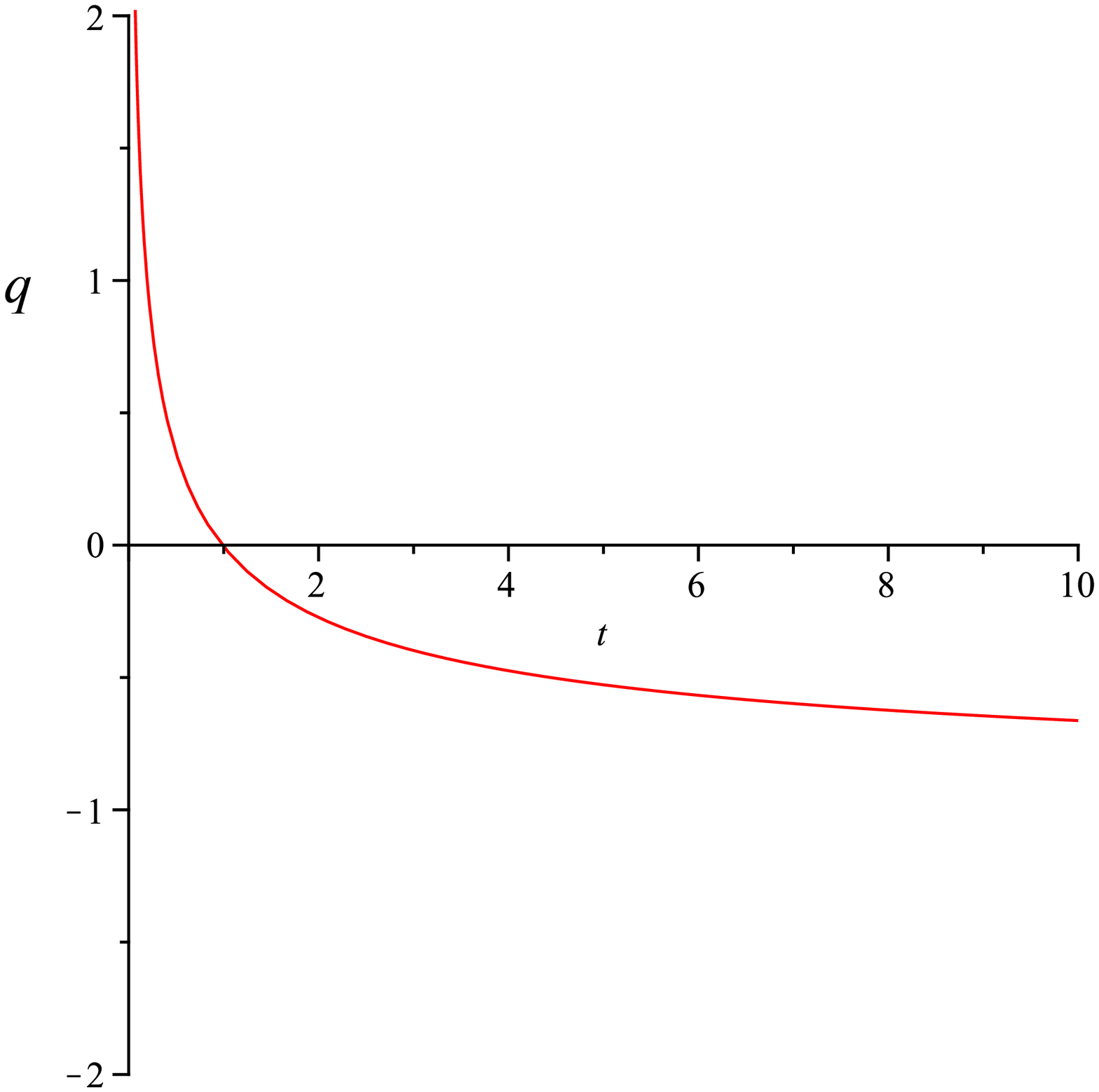}
     {{\caption{Plot of $a(\bar t)$,  $\psi(\bar t)$ and $q(\bar t)$ for the solutions (\ref{scalef.I}) and (\ref{dilaton.I}), with lower signs. The parameter
values  are $Q_{0} = -10$, $f_0 = -10$, $\bar y_{0} = 0$,  and $n=-4/3$.}}}
  \label{fig:1}
\end{figure}

 \subsection{$X(\bar x,\dot{\bar x})\sim \dot{\bar x} g(\bar x)$}

According to \cite{Hoj1}, we consider another ansatz $X(\bar x,\dot{\bar x})\sim \dot{\bar x} g(\bar x)$, where $ g(\bar x)$ is an arbitrary function. By this assumption, in order for $X$ to be the symmetry vector,  we obtain
 \begin{equation}\label{ff}
   f(\bar x)={{\frac{1}{2}}}\bar{\phi}'\,^2(\bar x)= \frac{
   g''(\bar x)}{ g'(\bar x)}>0.
   \end{equation}
By solving this equation we find  $\bar \phi$ in terms of $\bar x$ as

\begin{equation}\label{phix}
       \bar \phi(\bar x)=\bar\phi_c \pm{{ \sqrt{2}}}\int\sqrt{\frac{g''(\bar x)}{g'(\bar x)}}\mathrm{d}\bar x,
\end{equation}  
where  $\bar{\phi_c}$  is  constant of integration. Also, the Hojman conserved quantity is obtained as
\begin{equation}\label{Q_0}
\dot{\bar x}g'(\bar x)=Q_0.
\end{equation}
On the other hand, using equations (\ref{ff}) and (\ref{zdotx}) gives rise to

\begin{equation}\label{barzbarx}
        \bar{Z}(\bar x)=\bar{Z}_0 \frac{g''(\bar x)-3g'(\bar x)}{g'^3(\bar x)},
\end{equation}
which can be rewritten as
\begin{equation}\label{zbarx}
        \bar{Z}(\bar x)=\bar{Z}_0 (f-3)e^{-2\int_{\bar x_0}^{\bar x}f(\bar x')\mathrm{d}\bar x'}.
\end{equation}
{{Putting (\ref{barzbarx}) in (\ref{phibar}) and considering
(\ref{Q_0}), we find $\bar{Z}_0=-Q_0^2$ for which the solutions satisfy the field equations.}}
Now, by considering some ansatzs for $g(\bar x)$, we find some exact solutions
in the following.\vspace{3mm}

\begin{center}
        \bf{i. $g(\bar x)=\beta \, e^{\frac{1}{2} \alpha^2 \bar x}$}~~({\it{$\alpha$,
        $\beta$ are constants}})\\
\end{center}
By this choice,  Eq.(\ref{phix}) becomes
 \begin{equation}\label{phix1}
        \bar{\phi}(\bar x)=\bar{\phi_c}\pm{{\alpha}}(\bar x-\bar x_0),
 \end{equation}
 where $\bar x_0$ is constant of integration. Also, using the equations (\ref{Q_0}), (\ref{zbarx}) and (\ref{phix1}) yields
  
 \begin{equation}\label{barx}
        \bar x(\bar t)=\bar x_0+\frac{2}{\alpha^2}\ln \left( 1+e^{-\frac{1}{2}\alpha^2\bar x_0}\frac{Q_0}{\beta}(\bar t-\bar t_0) \right), 
 \end{equation}
 and
 
 \begin{equation}
        \bar{Z}(\bar{\phi})={{Q_0^2}}\:\frac{2({{6-\alpha^2}})}{\beta^2\alpha^4}e^{{\mp\alpha(\bar{\phi}-\bar{\phi_c})}}.
 \end{equation}
  Finally, by using the conformal transformations, we find \footnote{
{ By a time reparametrization as $t=1+\tau$ and absorbing
the parameters ${Q_0}, {\beta}$, and $t_0$ in this redefinition, accompanied
by re-scaling of scale factor through absorbing the parameter $\bar x_0$ in this redefinition, the above solutions can be rewritten simply as follows 
$$
a(t)={\sqrt{2}}\: t^{\frac{2\mp{\sqrt{2}}\alpha}{\alpha^2}},
    $$
    $$
       \psi(t)={2}\ln \left( \frac{1}{2}\,  t^{\pm \frac{{\sqrt{2}}}{\alpha}}\right),
   $$
where only $\alpha$ is left as the free parameter.}}
\begin{equation}\label{first g(x) a(t)}
a(\tau)={\sqrt{2}} e^{\bar x_0} \left(1+\tau\right)^ \frac{2{\mp} {\sqrt{2}}\alpha}{\alpha ^2},
    \end{equation}
    \begin{equation}\label{first g(x) sy(t)}
       \psi(\tau)={{2}}\ln \left [ \frac{1}{2}\, \left(1+\tau\right)^{{\pm \frac{{\sqrt{2}}}{\alpha}}}\right],
    \end{equation}
and
    \begin{equation}\label{first g(x) sy(t)w(sy)}
      ^{^{(2)}}W(\psi)={{Q_0^2}}{\frac{({{6-\alpha^2}})2^{(3\mp \alpha \sqrt{2})}}{\beta^2\alpha^4}\,e^{(1\mp \frac{\sqrt{2}}{2}\alpha)\psi}+\frac{1}{2}}\,,
     \end{equation}
where $\tau=\frac{Q_0}{\beta}e^{- \frac{1}{2}\alpha^2 \bar x_0} (\bar t-\bar t_0) $\,, for simplicity. {{Apart
from the integration constant $\bar{t}_0$ which can be absorbed in the time
reparametrization,  the solutions
depend implicitly on the conserved charge
$Q_0$ as well as the constant parameters, namely $\beta$, $\bar x_0$ and $\alpha$,  related to the conserved charge
$Q_0$ through Eqs.(\ref{Q_0}), (\ref{barx}) and $g(\bar x)$. 
}}      

{{In  Figure 2, the diagrams of $a(\bar t)$,  $\psi(\bar t)$ and $q(\bar t)$ are depicted for the solutions (\ref{first g(x) a(t)}) and (\ref{first g(x) sy(t)}), with lower signs, for  typical values of the parameters $Q_0$, $\bar x_0$, ${\alpha}$ and ${\beta}$.
This solution describes a universe which
 experiences an ever accelerating
de Sitter expansion.}}      
\begin{figure}[ht]
  \centering
  \includegraphics[width=2in]{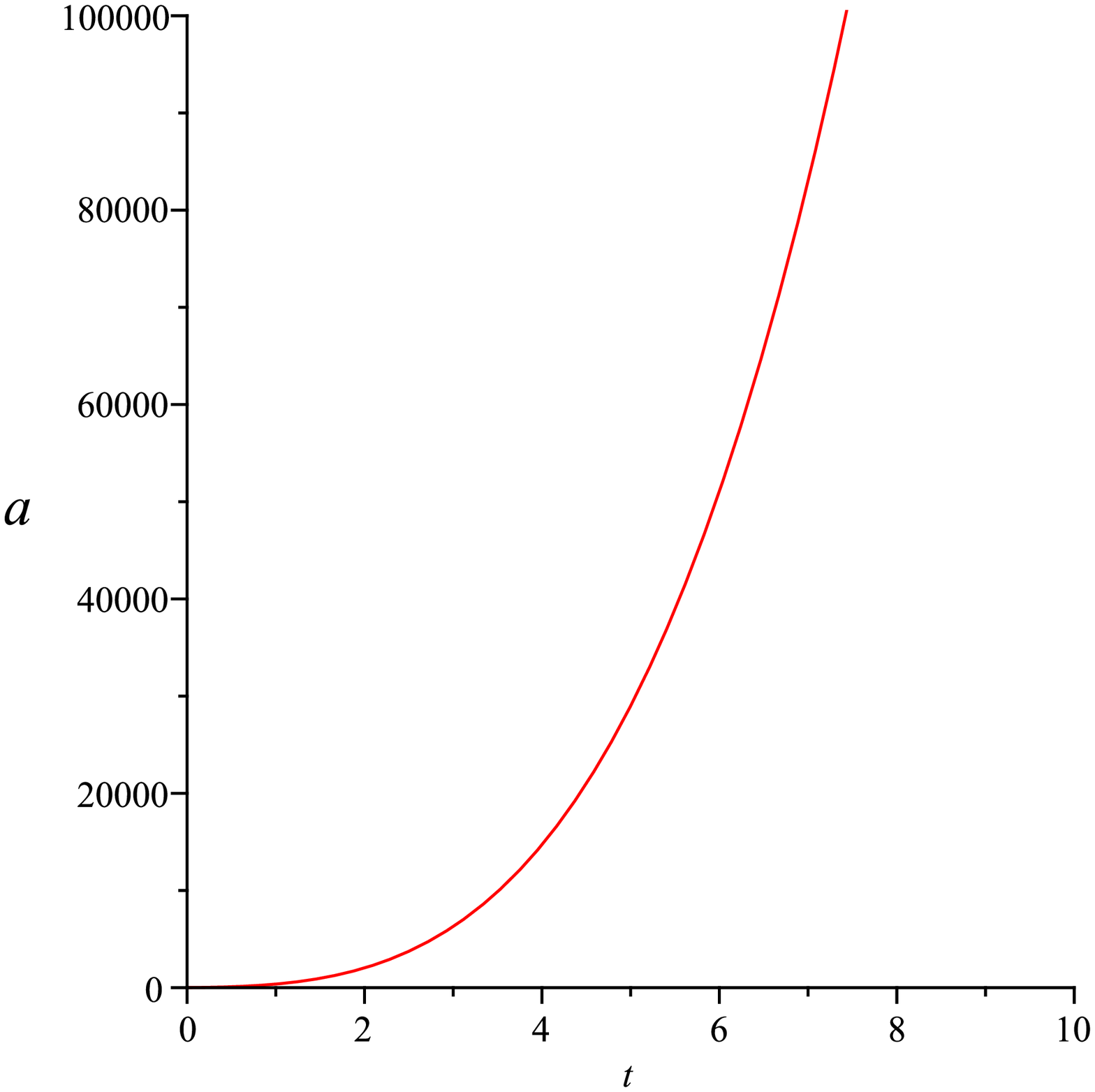}
  \includegraphics[width=2in]{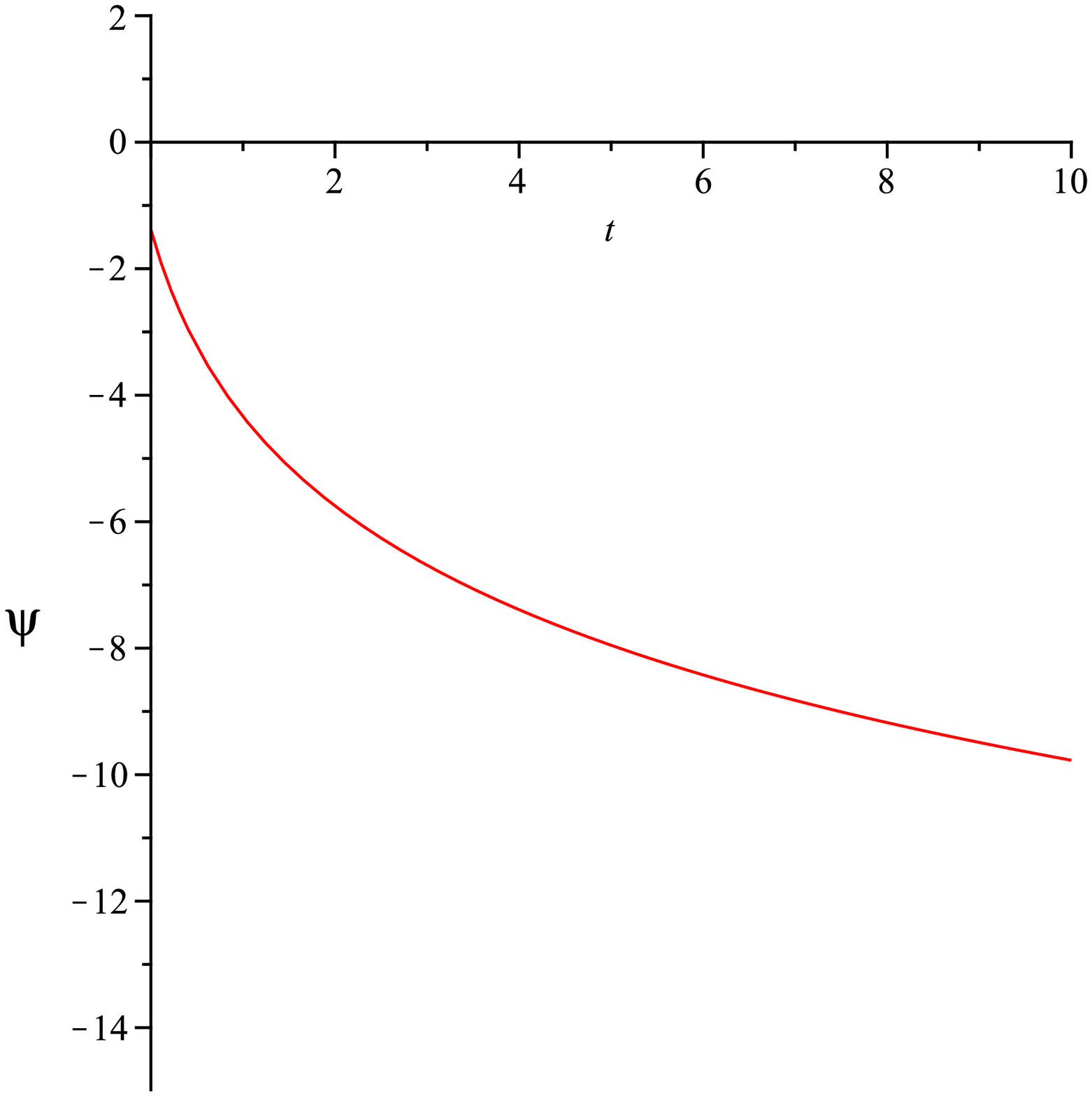}
  \includegraphics[width=2in]{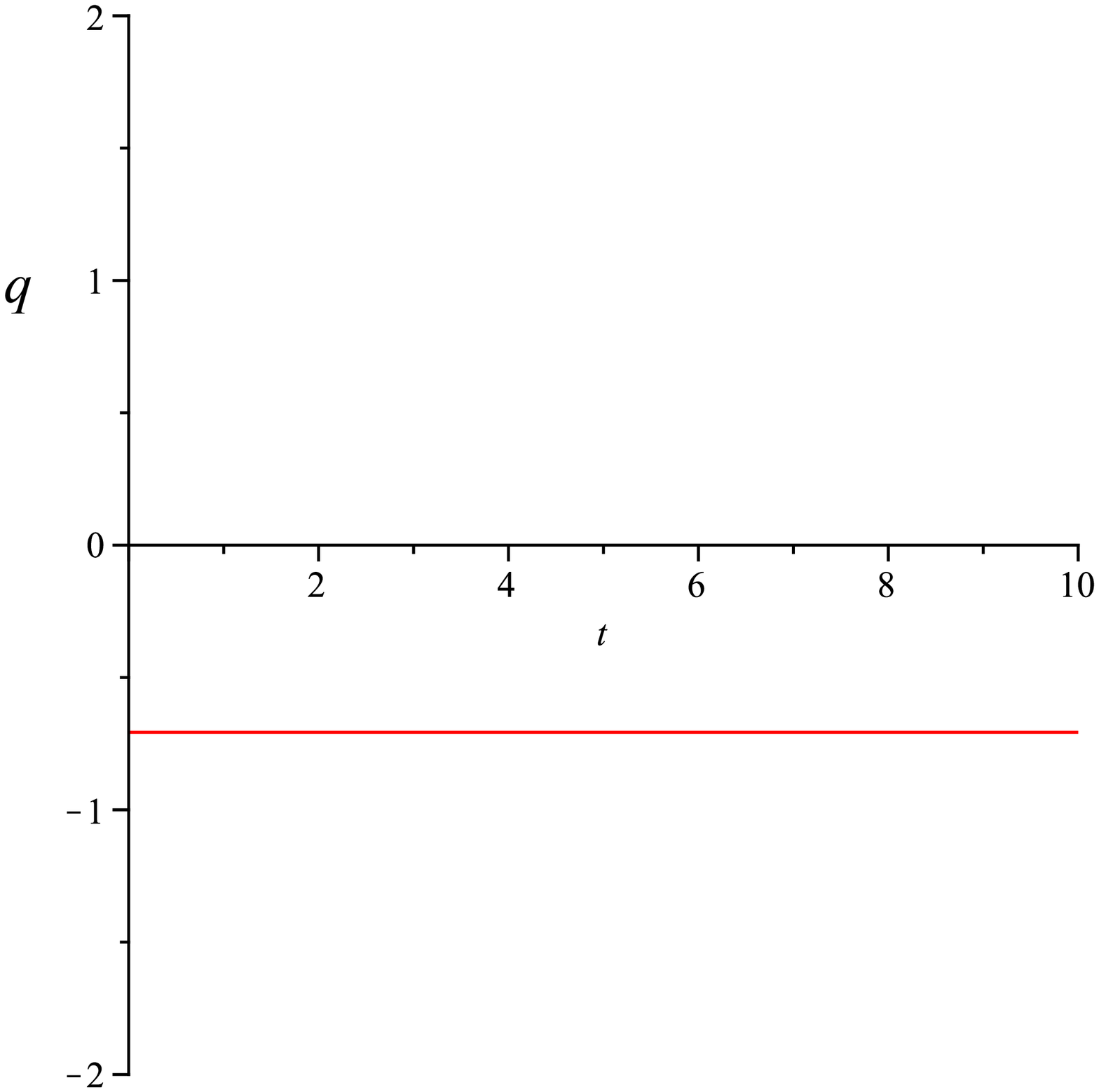} 
{{{ {\caption{Plot of $a(\bar t)$,  $\psi(\bar t)$ and $q(\bar t)$ for the solutions (\ref{first g(x) a(t)}) and (\ref{first g(x) sy(t)}),  with lower signs. The parameter
values  are  $Q_0=10$, $\bar x_0=2$, ${\alpha}=1$ and ${\beta}=2$. }}}}} 
\label{fig:2}
\end{figure} 
\vspace{3mm}
\begin{center}
\bf{ii. $g(\bar x)=\frac{(f_0+\bar x)^{1+\alpha}}{1+\alpha},\qquad \alpha>0 $}\\
\end{center}
Here $f_0$ and $\alpha$ are constants parameters. Again, from Eqs. (\ref{phix}), (\ref{Q_0}), (\ref{zbarx}) and (\ref{phix1}) we find respectively  $\bar{\phi}(\bar x)$, $\bar x(\bar t)$ and $\bar{Z}(\bar{\phi})$ as follows

\begin{equation}
        \bar{\phi}(\bar x)=\bar{\phi}_c \pm {{2\sqrt{2\alpha}}}\left[\left( \sqrt{f_0+\bar x}\right) \mp \left( \sqrt{f_0+\bar x_0}\right)\right], 
\end{equation}

\begin{equation}\label{xbart}
        \bar x(\bar t)=-f_0+\left[ (\alpha+1)Q_0(\bar t-\bar t_0)+(f_0+\bar x_0)^{\alpha+1}\right] ^{\frac{1}{1+\alpha}},
\end{equation}
and
\begin{equation}
        \bar{Z}(\bar{\phi})=\lambda \bar\Phi^{-4\alpha}- {{\frac{8\alpha^2}{3}}}\lambda \bar \Phi^{-4\alpha-2},
\end{equation}
where $\Phi=\bar{\phi}(\bar x)-\bar{\phi_c} \pm {2\sqrt{2\alpha}}\left( \sqrt{f_0+\bar x_0}\right) $ {{and ${\lambda={3Q_0^2}({8}\alpha)^{2\alpha}}$.}} Finally, using the conformal transformations yields the scale factor, the scalar field and the generic potential, respectively  as follows
\begin{equation}\label{second g(x)}
      a(\tau)=\sqrt{2}\, e^{-f_0}\, {e^{\pm 2\sqrt{\alpha }\sqrt{f_0+\bar x_0}}} \, e^{\mp 2 \sqrt{\alpha}((f_0+ \bar x_0)^{1+\alpha}+\tau)^ \frac{1}{2(1+\alpha)}}\, e^{((f_0+\bar x_0)^{1+\alpha}+\tau)^ \frac{1}{1+\alpha}},
\end{equation}
    \begin{equation}\label{sy2x}
      \psi(\tau)=\pm 4\sqrt{\alpha}\left((f_0+\bar x_0)^{1+\alpha}+\tau\right)^\frac{1}{2(1+\alpha)}+{\psi_0},
    \end{equation}
    and
    \begin{equation}\label{w2x}
      ^{^{(3)}}W(\psi)={4\lambda 2^{2\alpha}e^\psi \left[(\psi-\psi_0) ^{-4\alpha}-\frac{16}{3} \alpha^2(\psi-\psi_0) ^{-4\alpha-2}\right]}+\frac{1}{2} ,
    \end{equation}
    where $\tau=(1+\alpha)\, Q_0 \, (\bar t-\bar t_0)$ {and} ${\psi_0=\mp4\sqrt{\alpha}\sqrt{f_0+\bar x_0}-2\ln 2}$. {{Apart
from the integration constant $\bar{t}_0$ which can be absorbed in the time
reparametrization,  the solutions
depend implicitly on the conserved charge
$Q_0$ as well as the constant parameters, namely $f_0$, $\bar x_0$ and $\alpha$,  related to the conserved charge
$Q_0$ through Eqs.(\ref{Q_0}), (\ref{xbart}) and $g(\bar x)$. }}

\begin{figure}[ht]
  \centering
  \includegraphics[width=2in]{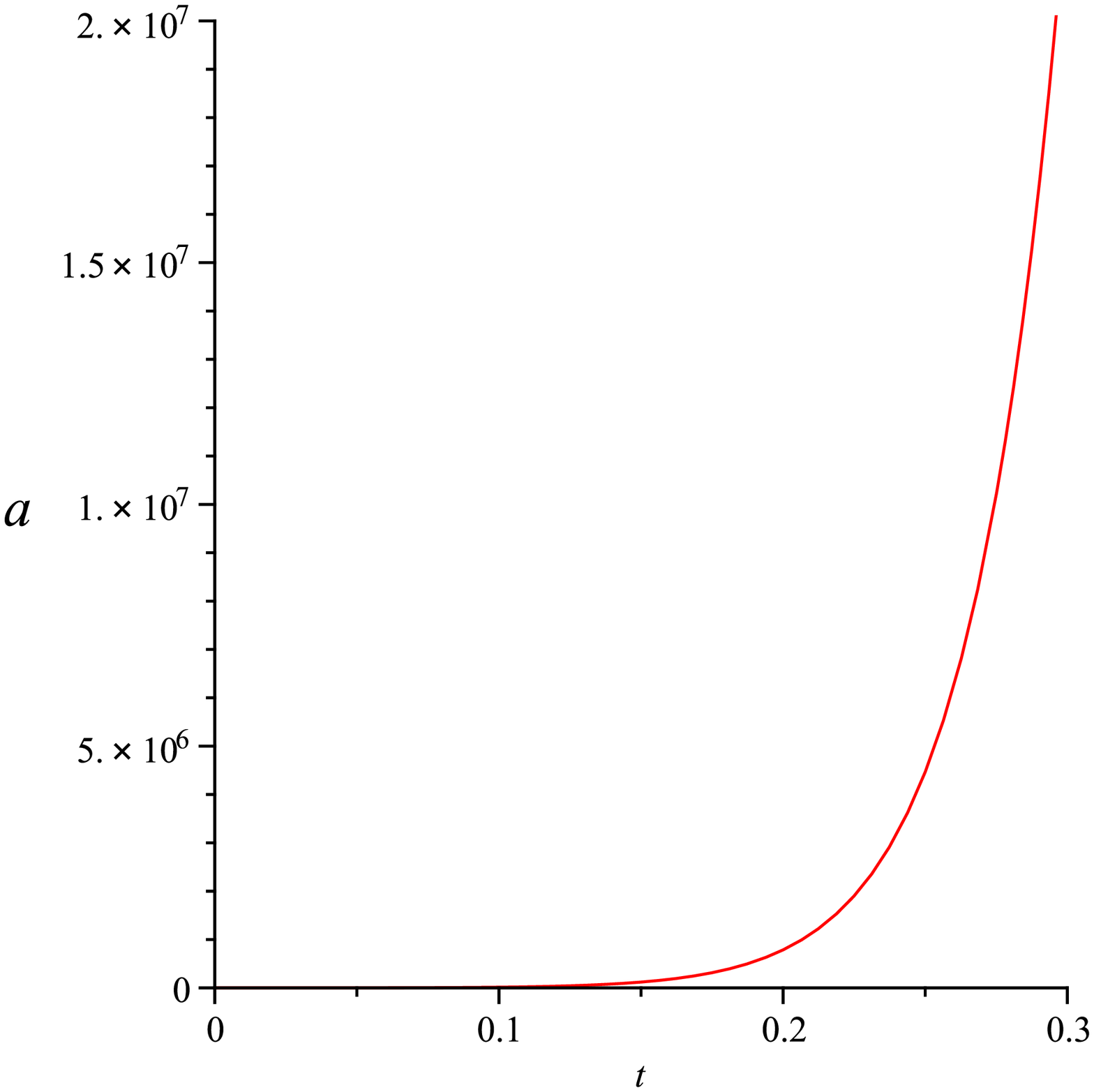}~~~~~
   \includegraphics[width=2in]{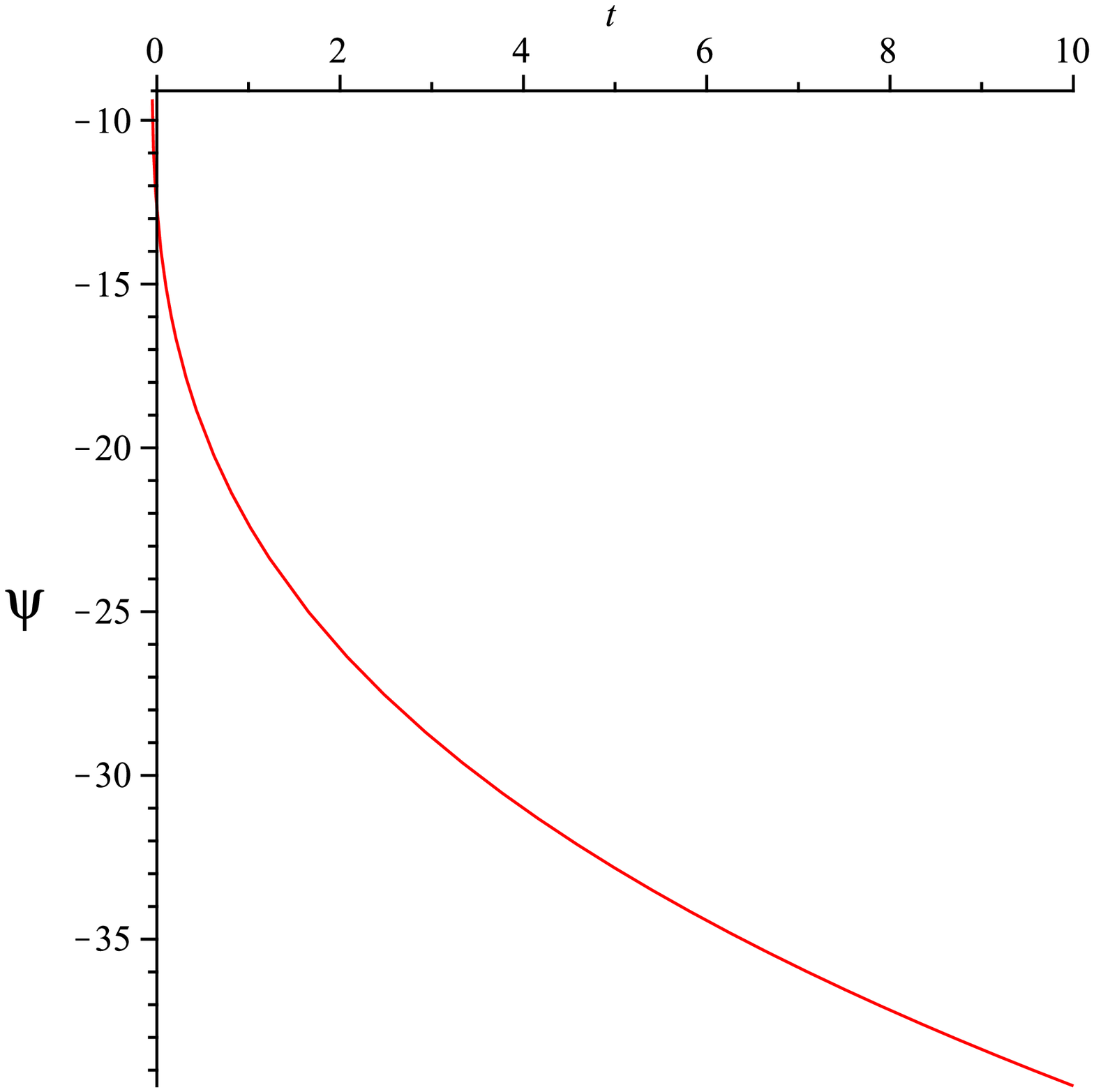}~~~~~
    \includegraphics[width=2in]{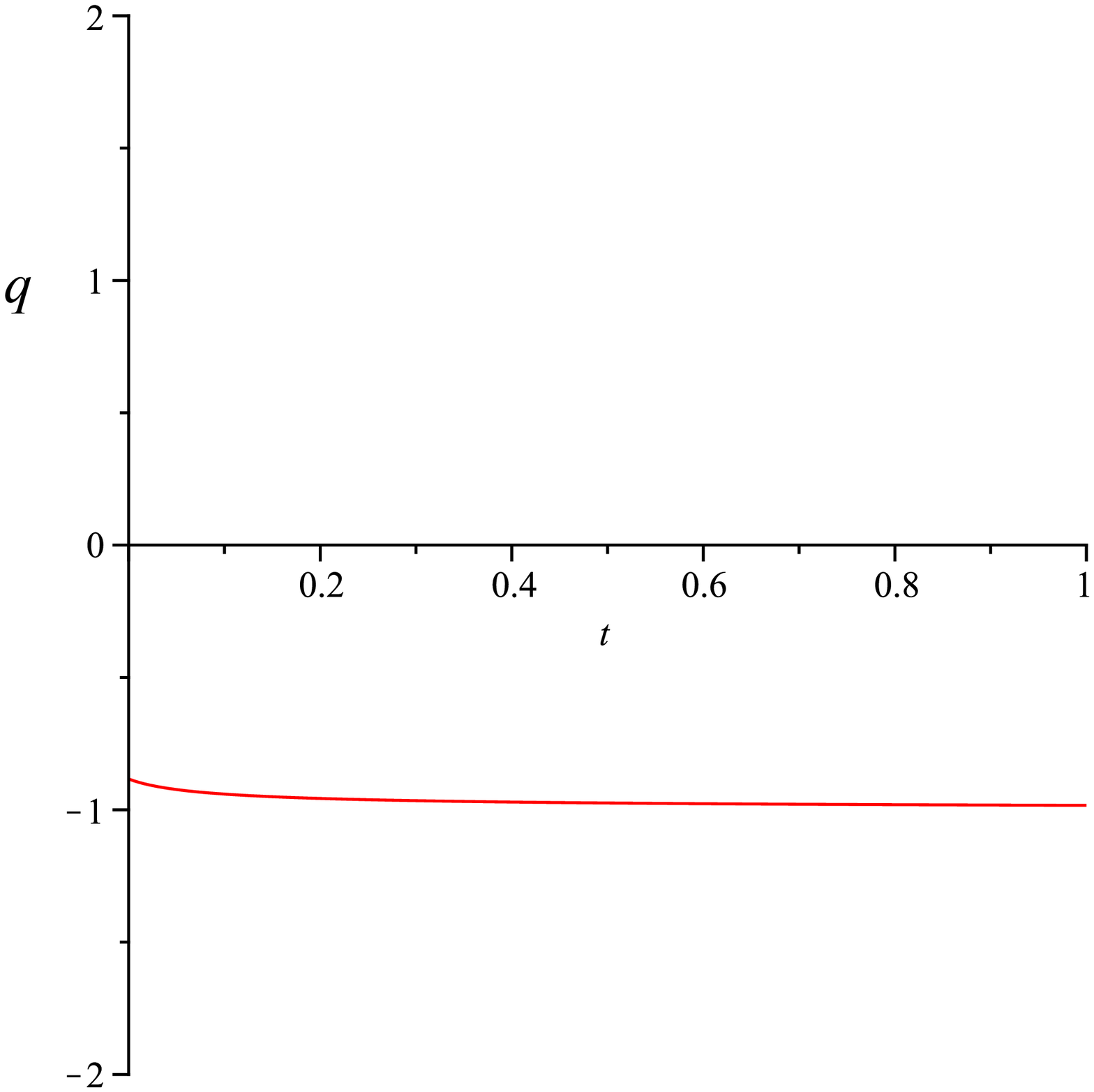}
     {{\caption{Plot of $a(\bar t)$,  $\psi(\bar t)$ and  $q(\bar t)$ for the solutions (\ref{second g(x)}) and (\ref{sy2x}), with lower signs. The parameter values  are $Q_{0} = 100$, $\bar x_0 = 4$, $f_0=0$, $\bar t_0=0$ and $\alpha=1/2 $.}}}
\label{fig:3}
\end{figure}

{{In Figure 3, the diagrams of $a(\bar t)$, $\psi(\bar t)$ and
$q(\bar t)$ are depicted for the solutions (\ref{second g(x)}) and (\ref{sy2x}), with lower signs, for some given values of the parameters $Q_0$,  ${\bar x_0}$, $f_0$, $\bar t_0$ and ${\alpha}$.  This solution describes a universe which experiences an ever accelerating de Sitter expansion.}}
\vspace{10mm}
\begin{center}
        \bf{ iii. $g(\bar x)=-\sqrt{2} \, \alpha \, (\bar x+e^{2\alpha^2 \bar x})$}\\
\end{center}
Here $\alpha$ is a constant parameter. Analogous to the previous cases, 

\begin{equation}
        \bar{\phi}(\bar x)=\bar{\phi_c} \pm {\frac{2}{\alpha}}\arcsinh(\sqrt{2} \,\alpha \, e^{\alpha^2 \bar x})\mp {\frac{2}{\alpha}\arcsinh(\sqrt{2} \,\alpha \, e^{\alpha^2 \bar x_0})},
\end{equation}
\begin{equation}\label{xbartau}
       \bar x (\tau)=\frac{1}{2\alpha^2}\left( \tau-\LambertW(2\alpha^2 e^\tau)\right), \end{equation}  
\begin{equation}
 \bar{Z}(\bar{\Phi})={{{\frac{Q_0^2}{2\alpha^2}}}}\left[2\alpha^2 \sech^6(\frac{1}{2}\alpha\bar{\Phi}) -(2\alpha^2-3) \sech^4(\frac{1}{2}\alpha\bar{\Phi})\right], 
\end{equation}
where ${\bar{\phi_c}}$ is a constant parameter, $\bar{\Phi}=\pm \frac{\sqrt{2}}{2}(\bar{\phi}-\bar{\phi_c}){+\frac{2}{\alpha}\arcsinh(\sqrt{2}\alpha e^{\alpha^2\bar x_0})}$ and $\tau=2\alpha^2Q_0(\bar t-\bar t_0)$. Also, using the conformal transformations here again we find
\begin{equation}\label{a(tau)}
        a(\tau)={\sqrt{2}e^{\pm \frac{\sqrt{2}}{\alpha}\arcsinh(\sqrt{2}\alpha e^{\alpha^2\bar x_0})}}\,{e^{\mp \frac{\sqrt{2}}{\alpha}\arcsinh\left( \sqrt{2} \alpha e^{\frac{1}{2}\left( \tau-\LambertW(2\alpha^2e^\tau)\right) }\right) }} e^{\frac{1}{2\alpha^2}\left( \tau-\LambertW(2\alpha^2e^\tau)\right) },
\end{equation}

\begin{equation}\label{psi(tau)}
\psi(\tau)=\psi_0\pm\frac{2\sqrt{2}}{\alpha}\arcsinh\left(\sqrt{2}\alpha e^{\frac{1}{2}\left(  \tau-\LambertW(2\alpha^2 e^\tau)\right) }\right), 
\end{equation}
\begin{equation}\label{W4}
^{^{(4)}}W(\psi)={{\frac{2Q_0^2}{\alpha^2}}}e^\psi\left[{\mp} (2\alpha^2-3)\sech^4\left( {\frac{\alpha}{4}(\psi+\psi_0 )}\right){\pm}2\alpha^2 \sech^6\left( {\frac{\alpha}{4}(\psi+\psi_0 )}\right) \right]+{\frac{1}{2}},
\end{equation}
where { $\psi_0=-2 \ln 2\mp \frac{2\sqrt{2}}{\alpha}\arcsinh(\sqrt{2}\alpha e^{2\alpha^2\bar x_0})$} is constant parameter. {{Apart
from the integration constant $\bar{t}_0$ which can be absorbed in the time
reparametrization,  the solutions
depend implicitly on the conserved charge
$Q_0$ as well as the constant parameters, namely $\bar x_0$ and $\alpha$,  related to the conserved charge
$Q_0$ through Eqs.(\ref{Q_0}), (\ref{xbartau}) and $g(\bar x)$. }}

{{In Figure 4, the diagrams of $a(\bar t)$, $\psi(\bar t)$ and
  $q(\bar t)$ are depicted for the solutions  (\ref{a(tau)}) and (\ref{psi(tau)}) with lower signs, for some given values of the parameters $Q_0$,  ${\bar
x_0}$, 
$\bar t_0$ and ${\alpha}$. This solution describes a universe which  experiences an ever accelerating 
de Sitter expansion. }} 
\begin{figure}[ht]
  \centering
  \includegraphics[width=2in]{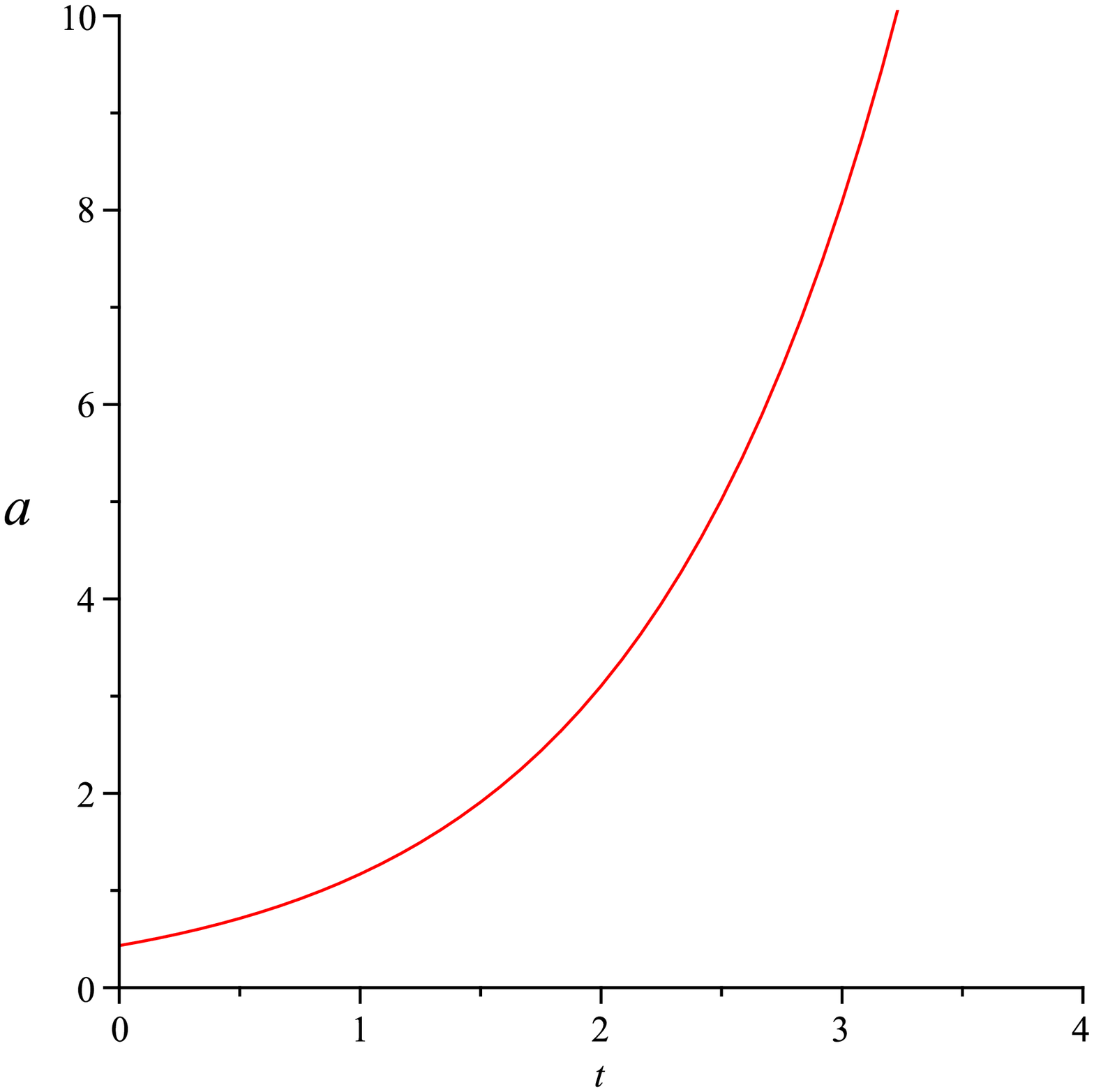}~~~~~
   \includegraphics[width=2in]{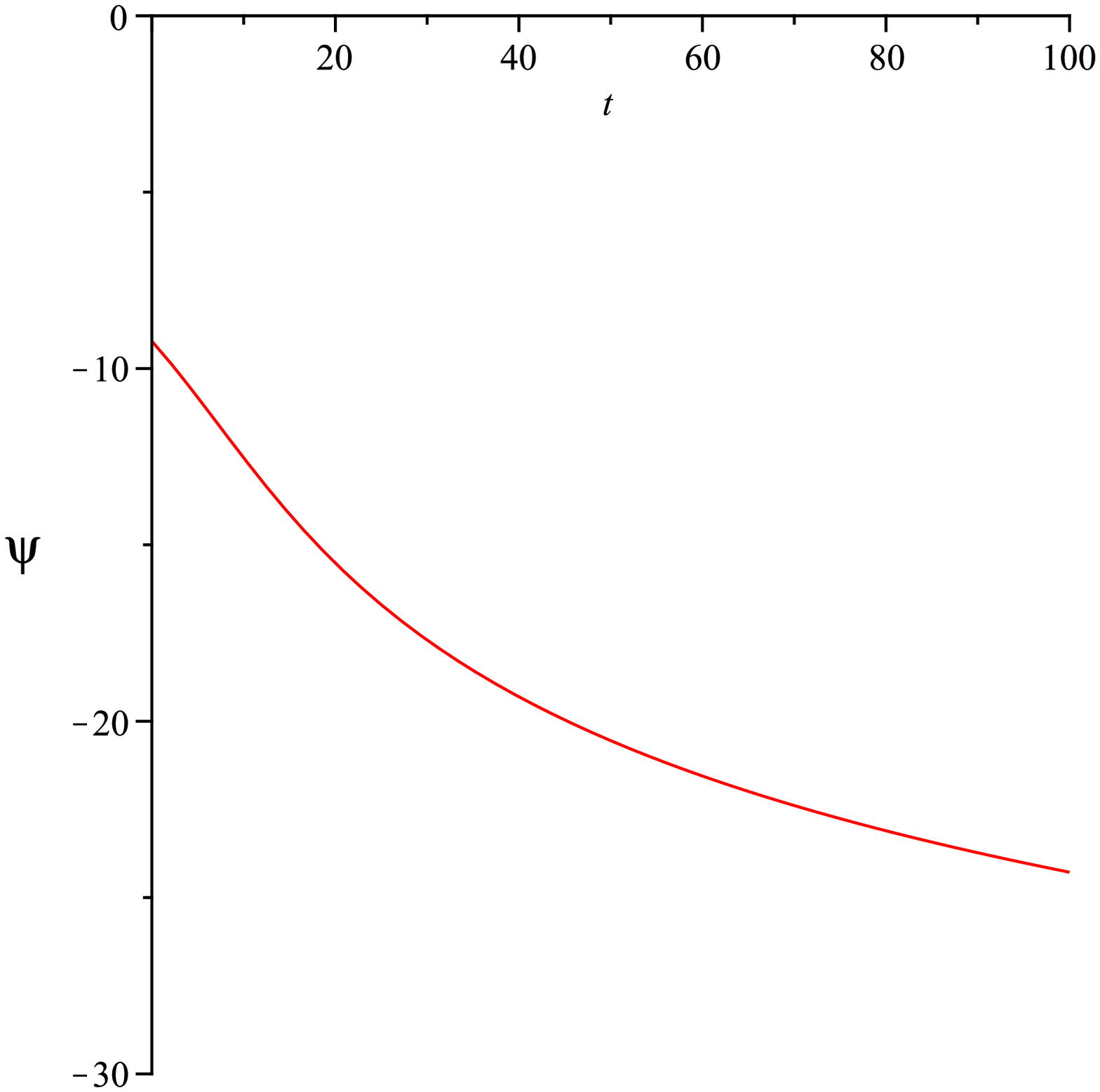}~~~~~
    \includegraphics[width=2in]{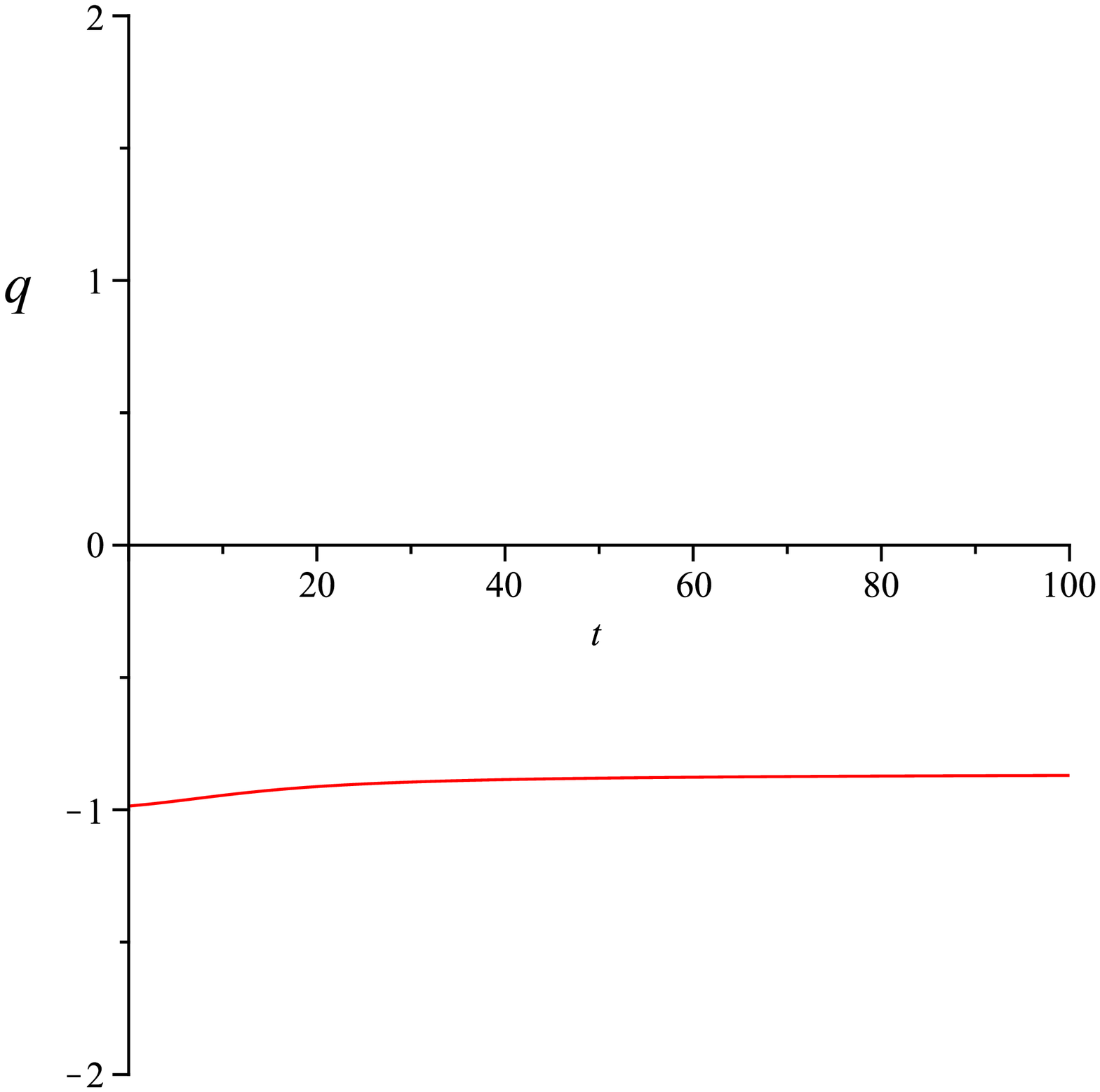}  
{ \caption{Plot of $a(\bar t)$,  $\psi(\bar t)$ and  $q(\bar t)$ for the solutions  (\ref{a(tau)}) and (\ref{psi(tau)}),  with lower signs. The parameter
values  are $Q_{0} = 1$, $\bar x_0 = 1$,  $\bar t_0=0$ and $\alpha=\sqrt{1/10} $.}}
\label{fig:4}
\end{figure}

{\section{Discussion}

We deem it necessary to discuss about  the characteristic features of obtained solutions and the corresponding
scalar field potentials, as follows. 

The first point is that there are a number of constants  
introduced in the solutions and the potentials, which can be divided into two sets. One set, like $\gamma_0$, $t_0$, $\bar
x_0$,
$y_0$,
$Z_0$, $\bar \varphi_c$, and $\bar \phi_c$ are just integration constants which
have no any specific meaning other than  specifying some initial values.
  The other set, like $Q_0$, $f_0$, $n$, $\alpha$ and $\beta$ are constants
  or parameters originating from Hojman symmetry. The suitable combinations of these
constants play key roles in the characteristic features of  obtained solutions and potentials. For instance, looking at the solutions (\ref{first g(x) a(t)})-(\ref{first g(x) sy(t)w(sy)}), we realize that they simply look like
the well-known power-law solutions of the form
$$
a(\tau)\sim \tau^p\:,\:\:\:\:\:\:\:\:\: \psi(\tau)\sim \ln (\tau^{q})\:,\:\:\:\:\:\:\:\:\: W(\psi)=\Lambda+Ae^{r\psi}\:,
$$
where $A$ is a constant, $\Lambda $ is a cosmological constant and the powers
$q$ and $r$ are related to the power $p$. It is worth to mention that the
physical characters of the solutions and the potentials are determined merely by the constants or parameters corresponding to Hojman symmetry. For instance,
$n$ and/or  $\alpha$ determine the time dependent behaviors of the solutions;
  $Q_0$, $\alpha$ and $\beta$ play role in determining the coupling constants
of the potentials; and  $n$ and/or $\alpha$  determine the $\psi$ dependence
of the potentials. The Hojman parameters $Q_0$, $n$, $\alpha$ and $\beta$
also play roles in the definition of time in the solutions, however, because
of the freedom in reparametrization of time, these roles are not of particular
importance.

The second point is that among the obtained typical solutions here, the first one corresponding to the potential $^{^{(1)}}W(\psi)$ exhibits an scale factor
behavior which changes from a deceleration to an acceleration phase, accounting for a {\it late time acceleration}, whereas  the solutions corresponding to other potentials $^{^{(2)}}W(\psi)$,
$^{^{(3)}}W(\psi)$, and $^{^{(4)}}W(\psi)$ show the scale factor behaviors
as they are experiencing {\it eternal inflationary expansions.} These ideal
accelerating
 behaviors
are clearly anticipated from the presence of a cosmological constant in the potentials,
however  including
some other realistic material elements in the original action may provide us with more realistic cosmological solutions. Moreover, we have obtained some limited solutions, using some  particular ansatz for the vector $X$ proposed in \cite{Hoj1}. A further study
may lead to other new and more realistic solutions, using some  other  ansatz for the vector $X$. }    

\section{Conclusion}

In this paper, we have studied an string cosmological model with the {FRW}
background metric accompanied by a totally antisymmetric field strength,
non-minimally coupled to a scalar field  
having a potential term. Using the Hojman symmetry
approach, we have fixed the scalar field potentials and obtained the corresponding
new exact solutions for the scale factor and the scalar field. {{The presence
of these free parameters, together with those of arising from Hojman symmetry,  is an important advantage using which
one can construct some cosmological  solutions $a(\bar t)$ and $\psi(\bar
t)$ with variety of cosmological behaviors. To evaluate the cosmological viability of the
solutions, we have depicted the
diagrams of $a(\bar t)$, $\psi(\bar t)$, and $q(\bar t)$, and discussed on their
 cosmic behaviors.
}}

\end{document}